\documentstyle[preprint,aps]{revtex}
\begin{document}
\title{\begin{flushright}
{\textnormal{IFIC/02-05}}
\vspace{1.5cm}
\end{flushright}
The Euler-Kockel-Heisenberg Lagrangian at Finite Temperature}
\author{J. L. Tomazelli \footnote{On leave from Departamento de F\'{\i}sica e 
Qu\'{\i}mica, UNESP, Guaratinguet\'a, Brazil.}}
\address{Departament de F\'{\i}sica Te\`orica, IFIC-CSIC, Universitat de Val\`encia \\
46100 Burjassot, Val\`encia, Spain.}
\author{L. C. Costa}
\address{Instituto de F\'{\i}sica Te\'{o}rica,\\
Universidade Estadual Paulista, \\
01405-900, S\~{a}o Paulo, Brazil.}
\maketitle

\begin{abstract}
In the present work we investigate temperature effects on the spinor 
and scalar effective QED, in the context of Thermo Field Dynamics. Following 
Weisskopf's zero-point energy method, the problem of charge renormalization 
is reexamined and high temperature contributions are extracted from the 
thermal correction for the Lagrangian densities.
\end{abstract}

%

\renewcommand{\thefootnote}{\fnsymbol{footnote}} %
%

%
%
\newpage

\section{Introduction}

The investigation of the finite temperature effects in Effective Quantum
Electrodynamics dates back the pioneering paper by W. Dittrich \cite{DI79}.
Motivated by the works of Weinberg \cite{WE74} and others \cite{DO74}, who
investigated finite temperature effects in a variety of models in quantum field theory,
Dittrich performed a detailed analysis of the finite temperature effects on the spinor 
and scalar QED in the presence of a constant magnetic field. In particular, he 
examined the Euler-Heisenberg effective Lagrangian \cite{EU36} at finite temperature 
in the context of the elegant Fock-Schwinger proper-time method \cite{SC51}.

Almost ten years ago, Loewe and Rojas \cite{LO92} extended Dittrich's work 
in the sense that they considered a general constant electromagnetic field. They 
found an effective thermal Euler-Heisenberg Lagrangian from which the effective 
thermal coupling constant and an effective thermal mass for the photon were identified. 
While Dittrich made use of the so-called imaginary time formalism \cite{MA55} in order 
to implement temperature effects, Loewe and Rojas employed the real-time formalism known
as Thermo Field Dynamics (TFD) \cite{UM84}. They argued that, despite the complication 
with doubling the number of fields, TFD allowed them to have a clear factorization of the 
interesting thermal effects. Along the same line, there are also the works by Elmfors 
{\it et al} \cite{EL93}, in which the problem of charged fermions in a (weak) constant 
magnetic field is investigated. They found that, in the high temperature limit, a 
logarithmic temperature dependence to the effective Lagrangian density is in agreement 
with \cite{LO92}.  

In another important work on the issue of nonlinear electromagnetic interactions at finite 
temperature, Brandt {\it et al} \cite{BR94} considered the analytically continued 
imaginary-time thermal perturbation theory \cite{BA90}, showing on quite general grounds 
that, for high temperatures, the electron-positron box diagram contribution to the 
effective action has a finite nonzero limit. In addition, from arguments of Lorentz 
invariance, they also addressed the absence of the logarithmic temperature dependence 
in the nonrelativistic static case as well as in the long wavelenght limit.

From this scenario, there remain many questions concerning the behaviour of the Effective 
QED in the (low and high) temperature limit and, therefore, further investigations must be 
done in order to elucidate some aparent contraditory results found in the literature.

With this in mind, the purpose of this paper is two fold: the first is to bring new light 
to the problem of high temperature effective spinor and scalar QED using an alternative 
approach based on TFD; the second, is to show the versatility of the (physicaly intuitive)
zero-point energy method due to Weisskopf \cite{WE36} which recently, has been improved 
from a mathematical standpoint in a paper by one of the authors \cite{JE00}. In 
\cite{JE00}, the (zero-temperature) Euler-Kockel-Heisenberg (E-K-H) Lagrangian was 
rederived for QED (spinor and scalar) and QED$_3$ following the approach due to 
Berestetskii, Lifshitz and Pitaeviski \cite{BE97} (which is inspired by Weisskopf ideas), employing Pauli-Villars-Rayski regularization 
prescrition \cite{RA48} in order to keeping track of divergent sums and momentum integrals.

Thus, following \cite{JE00}, we study, in the next section, the problem of charge 
renormalization and the high temperature contribuction is extracted from the thermal 
correction for the Lagrangian density. In section III, a similar calculation is 
done in the case of scalar QED and a closed expression for the high temperature 
E-K-H effective Lagrangian density is obtained. Finally, in section IV, we make 
some concluding remarks.
%
%
\section{The E-K-H Lagrangian at Finite Temperature}
\setcounter{footnote}{0}
The Dirac Hamiltonian for the electron-positron field in the presence of a
constant and homogeneous electromagnetic field is written in Fock space as
\begin{equation}
{\cal H} = \sum_{{\bf {p}}, \sigma} \epsilon_{{\bf {p}} \sigma} \; 
( a_{{\bf{p}} \sigma}^{\dagger} a_{{\bf {p}} \sigma} + 
b_{{\bf {p}} \sigma}^{\dagger} b_{{\bf {p}} \sigma} ) + \varepsilon_0 \;.
\end{equation}
where
\begin{equation}
\varepsilon_0 = \langle 0| {\cal H} |0 \rangle = - \sum_{{\bf {p}}, \sigma} 
\epsilon_{{\bf {p}} \sigma}^{(-)},
\end{equation}
is the vacuum zero-point energy and $\epsilon_{{\bf {p}} \sigma}^{(-)}$ is the 
energy density related to the negative frequency solution of Dirac equation in 
the presence of the electromagnetic field.

Following \cite{BE97}, the E-K-H effective Lagrangian density of the electromagnetic 
field, which accounts for the non-linear vacuum polarization effects is given by 
\begin{equation}
\delta {\cal L} = {\cal L}_{ren} - {\cal L}_0 = 
- [ \varepsilon_0 - (\varepsilon_0)_{E = H = 0} ],
\end{equation}
where ${\cal L}_0$ and ${\cal L}_{ren}$ are, respectively, the original and renormalized 
Lagrangian densities for the applied external electromagnetic field.

In order to study temperature effects in the E-K-H theory we make use of the 
so-called Thermo Field Dynamics (TFD). In TFD, temperature is introduced through a 
Bogoliubov transformation in the vacuum state of the system. In the present case, for each 
mode $({\bf p}, \sigma)$, the vacuum state of the electron-positron field in the external 
field transforms as
\begin{equation}
|0 \rangle \rightarrow |0 \rangle_{\beta} = (1 + {\rm e}^{- \beta
\epsilon})^{-1/2} \{ |0 \rangle_a \otimes |0\rangle_b + 
{\rm e}^{- \beta \epsilon / 2} a^{\dagger} {\tilde a} 
|0 \rangle_a \otimes b^{\dagger} {\tilde b} |0 \rangle_b \}.
\end{equation}
where ${\tilde a}$ and ${\tilde b}$ are annihilation operators of auxiliary fields which act 
on their corresponding vacuum sectors \cite{UM84} and $\beta = (1/k T)$ ($k$ is the Boltzmann 
constant and $T$ the temperature).

Considering the case where only a magnetic field is present, the natural thermal generalization 
of (2) is
\begin{eqnarray}
\langle 0| {\cal H} |0 \rangle_{\beta} &=& \varepsilon_0^{\beta}
= \sum_{{\bf {p}} \sigma} \varepsilon_{{\bf {p}} \sigma} 
\{ \langle 0 | a^{\dagger} a |0 \rangle_{\beta}
+  \langle 0 | b^{\dagger} b |0 \rangle_{\beta} 
- \langle 0 |1|0 \rangle_{\beta} \}  
\nonumber \\
&=& \sum_{{\bf {p}} \sigma} \epsilon_{{\bf {p}} \sigma} \left\{ 
\frac{1}{1 + {\rm e}^{\beta \epsilon_{{\bf {p}} \sigma}}} 
- \frac{1}{1 + {\rm e}^{- \beta \epsilon_{{\bf {p}} \sigma}}} 
\right\}
\end{eqnarray}
where
\begin{equation}
- \epsilon_{{\bf {p}}, \sigma}^{(-)} = - \sqrt{ m^2 + (2n + 1 - \sigma)
\vert e \vert H + p_z^2 }
\end{equation}
is the energy levels of an electron with charge $- \vert e \vert$ in a constant and 
uniform magnetic field $H_z = - H$. In (6) $n = 0, 1, 2, ...$ and $\sigma = \pm 1$. For 
$T \rightarrow 0$ the correct zero temperature result, expression (2), is recovered. 

It must be stressed that the above expression for the thermal zero-point energy was 
obtained from the TFD vacuum state (4), which, by construction, accounts for the 
interaction between the fermion field and a thermal reservoir. Hence, in constrast with 
the functional action approach, there is no physical reason to consider separately the 
zero temperature contribution (say the the E-K-H Lagrangian). In fact, this kind of 
factorization emerges in a natural way in the Weisskopf's formalism extended to finite 
temperature theories.

Considering the density of momentum states 
\begin{equation}
\frac {\vert e \vert H}{2 \pi} \frac{dp_z}{2 \pi},  \nonumber
\end{equation}
the sum over the $z$-component of the electron momenta in (5) goes to the continuum and 
the negative vacuum zero-point energy turns out to be 
\begin{equation}
\varepsilon_0^{\beta} = \frac{\vert e \vert H}{(2 \pi)^2} 
\sum_{n \sigma} \int_{-\infty}^{\infty} dp \; \epsilon_{n \sigma} \; 
\left\{ \frac{1}{1 + {\rm e}^{\beta \epsilon_{n \sigma}}} 
- \frac{1}{1 + {\rm e}^{- \beta \epsilon_{n \sigma}}} \right\}
\end{equation}
where we have already changed the notation $p_z \rightarrow p$. It is convenient to rewrite (8) as 
\begin{equation}
\varepsilon_0^{\beta} = - \frac{ \vert e \vert H}{(2 \pi)^2} 
\sum_{n \sigma} \int_{-\infty}^{\infty} dp \; \epsilon_{n \sigma} \; 
\left\{ 
\frac{ \sinh(\beta \epsilon_{n \sigma})} 
{1 + \cosh((\beta \epsilon_{n \sigma})}
\right\}.
\end{equation}
Despite the complication in solving the above divergent momentum integral one can investigate its 
high temperature limit in order to obtain an asymptotic expression for the thermal zero-point energy. 
Taking $\beta \rightarrow 0$ in (9) we find
\begin{equation}
\varepsilon_0^{\beta \rightarrow 0} = - \frac{ \vert e \vert H}{(2 \pi)^2}
\sum_{n \sigma} \int_{-\infty}^{\infty} dp \; \epsilon_{n \sigma} \; \left\{
\frac{ 6 \beta (\alpha_{n \sigma} + p^2)^{1/2}}{12 + \beta^2 (\alpha_{n \sigma} +
p^2)} \right\}
\end{equation}
where we have introduced $\alpha_{n, \sigma}$ through
\begin{equation}
\epsilon_{n \sigma}^2 = m^2 + (2n + 1 - \sigma) \vert e \vert H + p^2 =
\alpha_{n \sigma} + p^2 .  \nonumber
\end{equation}

Now, in order to keep under control the divergence in (10), we make use the gamma function integral 
representation, namely
\begin{equation}
\frac{1}{A^{1 + \delta}} = \frac{1}{\Gamma (1 + \delta)} \int_{0^+}^{\infty}
d\eta \; \eta^{\delta} {\rm e}^{- A \eta},
\end{equation}
valid for $\delta > - 1$, which allow us to make an analytical regularization \cite{SP68} through 
the choice of an appropriated $\delta$ value. Hence, it follows that
\begin{equation}
\varepsilon_0^{\beta \rightarrow 0} = - \frac{ \vert e \vert H}{(2 \pi)^2}
\frac{1}{\Gamma (1 + \delta)} \sum_{n \sigma} \int_{0^+}^{\infty} d \eta \;
\eta^{\delta} {\rm e}^{- (12 + \beta^2 \alpha_{n \sigma}) \eta} \left\{ 6
\sqrt{\pi} \alpha_{n \sigma} \; \eta^{-1/2} - 3 \eta^{-1} \frac{\partial}{\partial \beta}
\left( \sqrt{\frac{\pi}{\beta^{2} \eta}} \right) \right\},
\end{equation}
where the gaussian integration over $p$ has already been performed. Calculating the $\beta$ 
derivative in the last term of (13) and making use of the identity
\begin{eqnarray}
\sum_{n \sigma} {\rm e}^{- \beta^2 \alpha_{n \sigma} \eta} = 
{\rm e}^{- \beta^2 m^2 \eta} 
\left( 1 + 2 \sum_{n=1}^{\infty} 
{\rm e}^{- \beta^2 (2 \vert e \vert H n ) \eta } \right)  
= {\rm e}^{- \beta^2 m^2 \eta} \coth( \beta^2 \vert e \vert H \eta ),
\end{eqnarray}
we rewrite (13) as
\begin{equation}
\varepsilon_0^{\beta \rightarrow 0} = A + B + C
\end{equation}
where
\begin{equation}
A = - \frac{ \vert e \vert H}{(2 \pi)^2} 
\frac{6 \sqrt{\pi} m^2}{\Gamma (1 + \delta)} 
\int_{0^+}^{\infty} d \eta \; \eta^{- 3/2 + 1 + \delta} 
{\rm e}^{- ( 12 + \beta^2 m^2) \eta} \coth( \beta^2 \vert e \vert H \eta ),
\end{equation}
\begin{equation}
B = - \frac{ \vert e \vert H}{(2 \pi)^2} 
\frac{6 \sqrt{\pi} \vert e \vert H}{ \Gamma (1 + \delta)} 
\int_{0^+}^{\infty} d \eta \; \eta^{ - 3/2 + 1 + \delta} 
{\rm e}^{- ( 12 + \beta^2 m^2) \eta} {\rm csch}^2(\beta^2 \vert e \vert H \eta )
\end{equation}
and
\begin{equation}
C = \frac{ \vert e \vert H}{(2 \pi)^2} 
\frac{3 \sqrt{\pi}}{\beta^2 \Gamma (1 + \delta)} 
\int_{0^+}^{\infty} d \eta \; \eta^{ - 3/2 + \delta } 
{\rm e}^{- ( 12 + \beta^2 m^2) \eta} \coth( \beta^2 \vert e \vert H \eta ).
\end{equation}
The above integrals are finite objects and therefore the integrands in (16)-(18) can be 
expanded in power series for $\beta m << 1$, within convergence domain which corresponds 
to the analytic continuation $\delta > -1$. In order to recover the original theory we 
take the limit $\delta \rightarrow 0$, after performing the $\eta$-integration. As a result 
we obtain, up to $\vert e \vert^4$,
\begin{equation}
A \simeq \frac{3 \sqrt{12} m^2}{\pi \beta^2} \left( 1 + \frac{1}{24} \beta^2 m^2 \right) - 
\frac{{\vert e \vert}^2 H^2}{4 (12)^{3/2} \pi} m^2 \beta^2 \left( 1 - \frac{3}{24} \beta^2 m^2 \right) + 
\frac{\vert e \vert^4 H^4}{12 (12)^{7/2} \pi} m^2 \beta^6 ,
\end{equation}
\begin{equation}
B \simeq - \frac{2 (12)^{3/2}}{\pi} \frac{1}{\beta^4} \left( 1 + \frac{3}{24} \beta^2 m^2 \right) 
+ \frac{{\vert e \vert}^2 H^2}{2 \sqrt{12} \pi} \left( 1 - \frac{1}{24} \beta^2 m^2 \right) - 
\frac{3 \vert e \vert^4 H^4 \beta^4}{40 (12)^{5/2} \pi} \left( 1 - \frac{5}{24} \beta^2 m^2 \right) ,
\end{equation}
and
\begin{equation}
C \simeq - \frac{(12)^{3/2}}{\pi} \frac{1}{\beta^4} \left( 1 + \frac{3}{24} \beta^2 m^2 \right) 
+ \frac{{\vert e \vert}^2 H^2}{4 \sqrt{12} \pi} \left( 1 - \frac{1}{24} \beta^2 m^2 \right) - 
\frac{\vert e \vert^4 H^4 \beta^4}{80 (12)^{5/2} \pi} \left( 1 - \frac{5}{24} \beta^2 m^2 \right) ,
\end{equation}
so that (3) can be written as
\begin{eqnarray}
\delta {\cal L}^{\beta \rightarrow 0} &=& 
{\cal L}_{ren}^{\beta \rightarrow 0} - {\cal L}_0 = 
- [ \varepsilon_0^{\beta \rightarrow 0} 
- (\varepsilon_0)_{H = 0}^{\beta \rightarrow 0} ]  \nonumber \\
&\simeq& - \frac{{\vert e \vert}^2 H^2}{8 \pi} \left( \frac{6}{\sqrt{12}} 
- \frac{5}{(12)^{3/2}} m^2 \beta^2 \right) + 
\frac{\vert e \vert^4 H^4}{\pi} \left( \alpha \frac{m^2 \beta^6}{(12)^{7/2}} - \frac{7 \beta^4}{80 (12)^{5/2}} \right)
\end{eqnarray}
and, therefore, 
\begin{eqnarray}
{\cal L}_{ren}^{\beta \rightarrow 0} &=& {\cal L}_0 
+ \delta {\cal L}^{\beta\rightarrow 0}  \nonumber \\
&\simeq& - \frac{{\vert e \vert}^2 H^2}{8 \pi} 
\left[ 1 + \left( \frac{6}{\sqrt{12}} -  \frac{5}{(12)^{3/2}} m^2 \beta^2 \right) \right]  \nonumber \\
&=& - \frac{H^2 e^2_{ren}}{8 \pi},
\end{eqnarray}
where terms up to $\beta^2$ have been discarded and the renormalized charge is given by
\begin{equation}
e \rightarrow e_{ren}^{\beta \rightarrow 0} = 
e \left[ 1 + \left( \frac{6}{\sqrt{12}} - \frac{5}{(12)^{3/2}} m^2 \beta^2  \right) \right]^{1/2}.
\end{equation}
%
%

\section{Scalar QED}

When spin effects become negligible, spinor QED reduces to scalar QED and a
quite different theory then arise. In s-QED, the Hamiltonian for the charged
boson field in the presence of an external electromagnetic field turns out to be
\begin{equation}
{\cal H}=
\sum_{{\bf {p}}}\epsilon _{{\bf {p}}}\;(a_{{\bf {p}}}^{\dagger} a_{{\bf {p}}} 
+ b_{{\bf {p}}}^{\dagger} b_{{\bf {p}}})-\varepsilon _{0}\;.
\end{equation}
where
\begin{equation}
\varepsilon _{0}= \langle 0|{\cal H}|0 \rangle = 
-\sum_{{\bf {p}}}\epsilon _{{\bf {p}}}^{(-)}\;{\dot{=}}
-\sum_{{\bf {p}}}\epsilon _{{\bf {p}}}.
\end{equation}
Using the corresponding TFD-vacuum state for each mode of the boson field
\begin{equation}
|0 \rangle \rightarrow |0 \rangle_{\beta }=
(1+{\rm e}^{-\beta \epsilon})^{1/2}{\rm exp} \left\{ 
-\frac{\beta \epsilon }{2}(a^{\dagger }{\tilde{a}}^{\dagger}
\otimes b^{\dagger }{\tilde{b}}^{\dagger })\right\} 
|0\rangle _{a}\otimes |0\rangle_{b}
\end{equation}
and, again, considering the case where there is just a magnetic field, (26) becomes
\begin{equation}
\varepsilon _{0}^{\beta }=\langle 0|{\cal H}|0\rangle _{\beta}=
\sum_{{\bf {p}}}\epsilon_{{\bf {p}}} 
\left\{ \frac{-2+{\rm e}^{\beta \epsilon _{{\bf {p}}}}}
{1-{\rm e}^{\beta \epsilon _{{\bf {p}}}}}\right\}
\end{equation}
where,
\begin{equation}
-\epsilon _{{\bf {p}}}=-\sqrt{m^{2}+2n|e|H+p^{2}}
\end{equation}
is the free spin degeneracy energy levels of the boson field with charge $-|e|$, in a 
constant and uniform magnetic field $H_{z}=-H$. 
Considering the corresponding density of momentum states, we have
\begin{equation}
\varepsilon _{0}^{\beta }=\frac{|e|H}{(2\pi )^{2}} 
\sum_{n=0}^{\infty} \int_{-\infty }^{\infty} dp \;\epsilon _{n} \left\{ 
\frac{-2+{\rm e}^{\beta \epsilon _{n}}}{1-{\rm e}^{\beta \epsilon _{n}}}
\right\} .
\end{equation}

As in the fermion case, the above integral is difficult to be solved analytically. 
Again, we are interested in its asymptotic behavior. At first, we take the low 
temperature limit. In this limit, it is easy to see that expression (30) leads to 
the expected zero temperature result. On the other hand, for $T\rightarrow \infty $ 
one can determine {\it a priori}, the high temperature effective Lagrangian density 
defined in (3), i.e.,
\begin{eqnarray}
-\varepsilon _{0}^{\beta \rightarrow 0} = - \frac{|e|H}{(2\pi )^{2}}
\sum_{n=0}^{\infty }\int_{-\infty }^{\infty} dp \left\{ 
\frac{1}{\beta }-\epsilon _{n}-\frac{\beta \epsilon _{n}^{2}}{2} \right\}
\end{eqnarray}
where terms up to second order in $\beta$ were neglected.

The above momentum integral presents different degrees of divergences at 
the ultraviolet, which are more severe than in the spinor case, damanding a 
consistent subtraction scheme. In order to extract finite results for the high 
temperature zero-point energy, we employ Pauli-Villars-Rayski regularization 
prescription \cite{RA48}. Following \cite{JE00}, where the motivation behind 
the use of such regularization scheme is discussed in detail, we substitute (31) 
by its regularized expression
\begin{equation}
-(\varepsilon _{0}^{R})^{\beta \rightarrow 0}= 
- \sum_{i}c_{i}(\varepsilon_{0,i})^{\beta \rightarrow 0},
\end{equation}
where
\begin{equation}
(\varepsilon _{0,i})^{\beta \rightarrow 0}=-\frac{|e|H}{(2\pi )^{2}}
\sum_{n=0}^{\infty }\int_{-\infty }^{\infty }dp\left\{ \frac{1}{\beta}
-\epsilon _{n,i}-\frac{\beta \epsilon _{n,i}^{2}}{2}\right\}.
\end{equation}
The linear, quadratic, cubic and logarithmic divergence appearing in (31) might be 
eliminated by imposing on the coefficients $c_i$'s the following conditions 
\begin{equation}
\sum_{i}^{N}c_{i} = 0, \;\;\;\;\;\;\;\;\;\; \sum_{i}^{N}c_{i}m_{i}^{2} = 0.
\end{equation}
In (34), $N$ is the total number of regulators and the coefficients are such that $c_{0}=1$ 
and $m_{0}=m$ is the bare mass of the boson field. 

From this scenario, the regularized high temperature effective Lagrangian density 
$\delta {\cal L}^{R}$ is constructed following the same steps done in the last section. 
However, despite the first term in (31) which do not contribute to $\delta {\cal L}^{R}$ 
in virtue of the first condition in (34), the remaining two terms may be handled with the 
help of (12), keeping under control spurious {\it finite} terms that violates the invariance 
of the theory under reflections $H \rightarrow - H$. This kind of odd-parity terms, which 
also appear in the zero-temperature theory, are subtracted by the addition of finite conterterms 
with opposite sign. As a final result we obtain
\begin{eqnarray}
\delta {\cal L} &=& \frac{m^4}{8 \pi^2} \int_{0^+}^{\infty} d \eta \; 
\frac{{\rm e}^{- \eta}}{\eta^{3}} \left\{ - \eta b \; \coth (\eta b) 
+ 1 - \frac{1}{3} b^2 \eta^2 \right\}  \nonumber \\
&+& \frac{ m^5 \beta \sqrt \pi}{8 \pi^2} \int_{0^+}^{\infty} d \eta \; 
\frac{{\rm e}^{- \eta}}{\eta^{7/2}} \left\{ - \eta b \; \coth (\eta b) 
+ 1 - \frac{1}{3} b^2 \eta^2 \right\},
\end{eqnarray}
where $b = \vert e \vert H / m^2 $, and a charge renormalization has been
performed with the help of conditions (34).

%
%

\section{Concluding Remarks}

In this paper we applied the formalism of Thermo Field Dynamics in order to study 
temperature effects in the scope of spinor and scalar effective QED. Improving 
Weisskopf's zero-point energy method we constructed a high temperature E-K-H 
Lagrangian density addressing the problem of charge renormalization at high temperature.

In the spinor case, we found that the renormalized coupling constant, expression (24), 
reaches a maximum slightly above its unrenormalized value as $\beta$ approaches zero. 
This result is in complete agreement with that of Brandt {\it et al} \cite{BR94} in 
the sense that at high temperatures a finite nonzero contribution to the effective 
Lagrangian density is expected. We have also explicitly shown that for the case of a 
weak external magnetic field no logarithmic temperature dependence occurs. We must 
point out that such a logarithmic behaviour might be expected in the low temperature 
limit \cite{JE00}, where it may be related to the ultraviolet divergence of the theory 
at zero temperature \cite{BR94}. 

Yet, we might argue that, in the high temperature regime, a logarithmic dependence 
would be obtained if we improperly define the effective lagrangian (3) as 
\begin{equation}
\delta {\cal L} = - [ \varepsilon_0^{\beta \rightarrow 0} - 
(\varepsilon_0^{\beta \rightarrow \infty})_{E = H = 0} ],
\end{equation}
but than its physical meaning becomes obscured and unjustifiable in the context of 
Weisskopf approach, where the quantum nature of the {\it thermal} vacuum is accounted 
for since the very beginning, and the corresponding nonlinear contributions to Maxwell's 
Lagrangian do not appear as corrections to zero temperature QED. Therefore, there is no 
need for a previous charge renormalization in the low temperature sector\cite{JE00} since 
it is absent from expansions (10) and (31).

In section III, scalar QED is considered and a similar calculation is performed. The 
divergences appearing in the asymptotic expression for the thermal zero-point energy 
(31) were regularized by means of the Pauli-Villars-Rayski subtraction scheme. Together 
with the specific conditions arising from different degrees of divergence in each term 
of (31), the requirement of parity invariance demanded the addition of finite 
counterterms so that parity invariance was retained. 

In addition to the academic character of the subject, a better understanding of thermal 
effects in the context of Effective QED is fundamental, since the range of aplicability 
is broad \cite{DI02}. Along these lines, we also conclude that Weisskopf's method in 
the context of TFD provides a powerfull tool for the investigation of problems associated 
with radiative corrections in finite temperature quantum field theory. 

%
%
\vspace{0.2cm}
\noindent {\bf Acknowledgements.} 
The authors would like to thank Holger Gies for helpful comments and references. 
JLT thanks Funda\c c\~ao de Amparo \`a Pesquisa  do Estado de S\~ao Paulo 
(FAPESP, Brazil) grant 2000/14758-2 and Spanish MCyT grant PB98-0693, for the 
partial financial support. LCC is grateful to FAPESP for the financial support.
%
\newpage

\end{document}